\documentclass[reprint,twocolumn,showpacs,showkeys,superscriptaddress,preprintnumbers,amsmath,amssymb,floatfix,aps,prb]{revtex4-1}

\usepackage{graphicx}
\usepackage{bm}

\usepackage{placeins}
\usepackage{graphics}
\usepackage{color}
\usepackage{hyperref}
\usepackage{multirow}

\begin{document}


\title{Giant proximity exchange and valley splitting in TMDC/hBN/(Co, Ni) heterostructures}

\author{Klaus Zollner}
\email{klaus.zollner@physik.uni-regensburg.de}
\affiliation{Institute for Theoretical Physics, University of Regensburg, 93040 Regensburg, Germany}

\author{Paulo E. Faria~Junior}
\affiliation{Institute for Theoretical Physics, University of Regensburg, 93040 Regensburg, Germany}

\author{Jaroslav Fabian}
\affiliation{Institute for Theoretical Physics, University of Regensburg, 93040 Regensburg, Germany}

\date{\today}

\begin{abstract}
We investigate the proximity-induced exchange coupling in transition-metal dichalcogenides (TMDCs),
originating from spin injector geometries composed of 
hexagonal boron-nitride (hBN) and ferromagnetic (FM) 
cobalt (Co) or nickel (Ni), from first-principles.
We employ a minimal tight-binding Hamiltonian that captures the low energy bands of the TMDCs 
around K and K' valleys, to extract orbital, spin-orbit, and exchange parameters.
The TMDC/hBN/FM heterostructure calculations show that due to the hBN buffer layer, 
the band structure of the TMDC is preserved, with an additional
proximity-induced exchange splitting in the bands.
We extract proximity exchange parameters in the 1--10 meV range, depending on the FM.
The combination of proximity-induced exchange
and intrinsic spin-orbit coupling (SOC) of the TMDCs, 
leads to a valley polarization, translating into 
magnetic exchange fields of tens of Tesla.
The extracted parameters are useful for subsequent exciton calculations 
of TMDCs in the presence of a hBN/FM spin injector. 
Our calculated absorption spectra show large splittings for the exciton peaks; in the case of
MoS$_2$/hBN/Co we find a value of about 8 meV, corresponding to about 50 Tesla external magnetic field
in bare TMDCs. 
The reason lies in the band structure, where a hybridization with Co $d$ orbitals causes a giant valence band exchange splitting of more than 10~meV. 
Structures with Ni do not show any $d$ level hybridization features, but still sizeable proximity exchange and exciton peak splittings of around 2~meV are present in the TMDCs.
\end{abstract}

\maketitle

\section{Introduction}
\label{sec:Introduction}
Spintronics is based on the efficient injection, transport, manipulation 
and detection of spins in a material \cite{Zutic2004:RMP, Han2014:NN, Fabian2007:APS}.
The current generation of spintronics devices employ hybrid geometries consisting of several
two-dimensional (2D) materials \cite{Liu2019:PMS, Briggs2019:2DM} in order to overcome 
intrinsic limitations of the transport medium. 
A new branch of physics has emerged, which is solely dedicated to the interface 
engineering \cite{Yuan2017:C, Hu2018:CSR} of those ultrathin layers, including
semiconductors, ferromagnets and superconductors, leading to new device technologies. 

The most prominent example in this field is graphene, 
which has intrinsically outstanding spin and charge transport properties. 
Since several years now, physicists study graphene based spintronics 
devices \cite{Han2014:NN} and have found an efficient way to inject spins
via FM/insulator tunnel junctions into 
graphene \cite{Kamalakar2014:SR, Fu2014:JAP, Drogeler2016:NL,Drogeler2014:NL, Kamalakar2016:SR, Kamalakar2015:NC}. 
Nonlocal measurement geometries reveal the spin transport properties of graphene, 
which can be modified by the presence of various 2D materials in van der Waals heterostructures.
Currently, state of the art spin transport geometries are based on 
hBN encapsulated graphene 
\cite{Drogeler2016:NL,Kamalakar2016:SR, Gurram2017:2DM, Gurram2017:NC}, where
spins are injected by FMs, with giant mobilities 
up to 10$^6$ cm$^2$/Vs \cite{Banszerus2015:SA, Petrone2012:NL, Calado2014:APL} 
and spin lifetimes exceeding 10~ns \cite{Drogeler2016:NL}. 
The insulating hBN is ideal to reduce the contact resistance to the FM 
and helps to preserve the linear dispersion of graphene, 
which is desired for spin transport \cite{Zollner2016:PRB}. Also oxide insulators are used, 
such as MgO and SiO$_2$, resulting in less efficient spin injection \cite{Gurram2017:2DM}.
The manipulation of spins can be achieved by inducing SOC or exchange coupling from proximity.
For example, a TMDC in proximity to graphene, 
induces strong valley Zeeman spin-orbit fields on the order of 1~meV \cite{Gmitra2016:PRB, Gmitra2015:PRB, Cummings2017:PRL}, 
significantly reducing spin lifetimes in graphene to about 10 ps \cite{Zihlmann2018:PRB, Omar2018:PRB, Avsar2017:ACS}, 
but leading to giant spin relaxation anisotropies (the ratio of out-of-plane to in-plane spin lifetime) of about 10 \cite{Cummings2017:PRL, Leutenantsmeyer2018:PRL, Xu2018:PRL, Zihlmann2018:PRB, Omar2018:PRB, Avsar2017:ACS}.
Additionally, it has been demonstrated that a TMDC can be 
utilized for optical spin injection in graphene \cite{Gmitra2015:PRB, Avsar2017:ACS, Luo2017:NL}, 
relevant for the field of optospintronics.

Recently, there has been a lot of effort to use also other 2D materials 
such as phosphorene or TMDCs as the transport medium.
The advantage is that they are already semiconducting, making them suitable 
for electronic and spintronic devices \cite{Zutic2004:RMP, Fabian2007:APS}
such as spin diodes and field effect transistors 
\cite{Li2014:NN, Zutic2006:IBM, Liang2017:JAP, Ahmed2017:NML}.
In the case of phosphorene, very little is known about its 
spin transport properties \cite{Kurpas2016:PRB, Kurpas2018:JPD, Avsar2017:NL};
the electronic ones are highly
anisotropic and show large mobilities \cite{Qiao2014:NC, Liu2014:ACS}.
Furthermore, measurements combined with first-principles calculations show, 
that hBN is also an ideal tunnel barrier,
when injecting spin polarized carriers from Co into phosphorene \cite{Avsar2017:NL}. 

Monolayer TMDCs have a band gap 
in the optical range and the valley degree of freedom plays a major role \cite{Kormanyos2014:2DM, Liu2015:CSR,Tonndorf2013:OE,Tongay2012:NL,Eda2011:NL,Xiao2012:PRL} in their (opto-) electronic properties.
The helicity selective excitation of carriers with 
certain spin in a certain valley at the same excitation energy, 
makes those materials very attractive for the field of valleytronics \cite{Vitale2018:S, Langer2018:Nat,Schaibley2016:NRM, Ye2016:NN, Zhong2017:SA}.
By an exchange field one can break the time-reversal symmetry of the TMDC, the degeneracy of the valleys, and introduce valley polarization.
In an external magnetic field, the valley splitting rises roughly linear with 0.1 -- 0.2 meV/T 
\cite{Srivastava2015:NP, Aivazian2015:NP, Li2014:PRL, MacNeill2015:PRL}, 
such that large fields are required to get a sizable effect.
A much better perspective to achieve large valley polarization in TMDCs 
is by the proximity exchange effect. 
Indeed, placing the TMDC on a magnetic substrate, giant valley splittings can be achieved, ranging from few to hundreds of meV 
\cite{Ji2018:PCCP, Li2018:PCCP, Qi2015:PRB, Zhang2016:AM, Xu2018:PRB, Zhao2017:NN, Ye2016:NN, Zhong2017:SA, Seyler2018:NL, Peng2017:ACS,Zhang2019:PRB}, which can additionally be tuned by gating and twisting \cite{Zollner2019a:PRB}.

For spin injection, one can either contact the TMDC with metal or metal/insulator interfaces or inject them optically \cite{Somvanshi2017:PRB, Sanchez2016:NL}. In the former case,
studying the dependence of the Schottky barrier on the used 
electrode is crucial \cite{Farmanbar2016:PRB, Farmanbar2015:PRB, Cui2017:NL, Zhao2017:AFM, Garandel2017:PRB, Rai2018:C, Schulman2018:CSR, Guo2014:ACS}. 
It turns out that a hBN tunnel barrier is also a good choice here, 
preserving the intrinsic properties of the TMDC while enormously reducing the contact resistance \cite{Farmanbar2015:PRB, Zhu2018:APL, Su2017:ACS, Wang2016:AM}.
Also other insulating barriers, such as TiO$_2$, MgO or Al$_2$O$_3$
\cite{Kaushik2016:ACS, Liang2017:NC, Wang2014:SR, Dankert2014:ACS, Hayakawa2018:JJAP}
are promising candidates, where the thickness of the barrier plays an important role
for the efficiency of spin injection.
The contact resistance can also be decreased by strong electron doping \cite{Khalil2015:ACS}
or using graphene electrodes \cite{Liu2015:NL}.

Electronic and spin transport \cite{Song2013:PRL, Tian2017:PRB, DuranRetamal2018:CS, Schmidt2015:CSR} in TMDCs is becoming an important topic.
It has been shown that the carrier mobility increases 
with the number of TMDC layers \cite{Rai2018:C, DuranRetamal2018:CS}, 
due to reduced Coulomb scattering in thicker samples \cite{Li2013:NL}, while phonon scattering 
limits the room temperature mobility \cite{Schmidt2015:CSR}.
Spin transport has been studied on a theoretical level \cite{Ochoa2013:PRB, Habe2016:PRB, Yang2015:PRB, Yang2016:PRB, Habe2016:PRB}, 
while spin injection has been demonstrated
electrically \cite{Ye2016:NN, Liang2017:NC} and optically \cite{Sanchez2016:NL}, showing spin diffusion lengths of about 200~nm in multilayer TMDCs \cite{Liang2017:NC}.

A very natural choice for spin injection and for
generating proximity exchange in TMDCs are hBN/FM tunnel structures. How large is the proximity-induced exchange in TMDC/hBN/FM heterostructures? What are the optical signatures of such structures? 
In this manuscript we study TMDC/hBN/FM tunneling spin injection heterostructures. 
We calculate the band
structure and employ a minimal tight-binding model Hamiltonian to extract orbital, 
spin-orbit, and proximity exchange parameters for the proximitized TMDCs, MoS$_2$, and WS$_2$.
Proximity exchange in the TMDCs is found to be on the order of 1--10 meV, 
and together with the intrinsic SOC of the TMDC, leads to a valley polarization
corresponding to tens of Tesla exchange field for bare TMDCs. 
Specifically the MoS$_2$/hBN/Co heterostructure shows a giant valence band spin splitting, 
of more than 10~meV, due to proximity exchange and hybridization 
of the TMDC valence band orbitals with Co $d$ orbitals. 
The corresponding calculated absorption spectrum shows a giant valley splitting of about 8~meV. 
The valley splitting for WS$_2$/hBN/Co is of similar magnitude (5~meV), despite the absence of 
hybridizing $d$ orbitals. In the case of Ni, proximity exchange and valley splittings are  
reduced (1--3~meV) for both TMDCs.
Our investigations should be useful for interpreting spin injection, spin 
tunneling and optical properties of TMDC/hBN/(Co, Ni) systems. Furthermore, 
the extracted parameters can be used for transport simulations 
and further studies of excitonic effects.

\section{Model Hamiltonian}
\label{sec:Hamiltonian}
As basis states for our model we use $|\Psi_{\textrm{CB}}\rangle = |d_{z^2}\rangle$ and 
$|\Psi_{\textrm{VB}}^{\tau}\rangle = \frac{1}{\sqrt{2}}(|d_{x^2-y^2}\rangle+\textrm{i}\tau |d_{xy}\rangle)$, 
corresponding to conduction band (CB) and valence band (VB) at K ($\tau~=~1$) and K' ($\tau~=~-1$),
since the band edges of bare TMDC monolayers are formed by different $d$-orbitals from the transiton metal \cite{Kormanyos2014:2DM}.
The model Hamiltonian, in the basis which includes electron spin $|\Psi_{\textrm{CB}}, \uparrow\rangle$, 
$|\Psi_{\textrm{VB}}^{\tau}, \uparrow\rangle$, $|\Psi_{\textrm{CB}}, \downarrow\rangle$, 
and $|\Psi_{\textrm{VB}}^{\tau}, \downarrow\rangle$, to describe 
the band structure of the TMDC close to K and K', in the presence
of proximity exchange \cite{Qi2015:PRB,Zollner2019a:PRB} is
\begin{flalign}
\label{Eq:Hamiltonian}
&\mathcal{H} = \mathcal{H}_{0}+\mathcal{H}_{\Delta}+\mathcal{H}_{\textrm{soc}}+\mathcal{H}_{\textrm{ex}}+\mathcal{H}_{\textrm{R}},\\
&\mathcal{H}_{0} = \hbar v_{\textrm{F}} s_0 \otimes (\tau\sigma_{x}k_{x}+\sigma_{y}k_{y}),\\
&\mathcal{H}_{\Delta} = \frac{\Delta}{2}s_{0}\otimes \sigma_{z},\\
&\mathcal{H}_{\textrm{soc}} = \tau s_{z} \otimes (\lambda_{\textrm{c}}\sigma_{+} + \lambda_{\textrm{v}} \sigma_{-}),\\
&\mathcal{H}_{\textrm{ex}} = -s_{z} \otimes (B_{\textrm{c}}\sigma_{+} + B_{\textrm{v}} \sigma_{-}),\\
&\mathcal{H}_{\textrm{R}} = \lambda_{\textrm{R}} (\tau s_{y}\otimes \sigma_{x} - s_{x}\otimes \sigma_{y}).
\end{flalign}
Here $v_{\textrm{F}}$ is the Fermi velocity. 
The Cartesian components $k_{x}$ and $k_{y}$ of the electron wave vector are measured from K~(K'). 
The pseudospin Pauli matrices are $\sigma_{\textrm{i}}$ acting on
the (CB,VB) subspace and spin Pauli matrices are $s_{\textrm{i}}$ 
acting on the ($\uparrow,\downarrow$) subspace,
with $\textrm{i} = \{0,x,y,z\}$. For shorter notation we introduce $ \sigma_{\pm} = \frac{1}{2}(\sigma_{0} \pm \sigma_{z})$.
TMDCs are semiconductors, and thus $\mathcal{H}_{\Delta}$ introduces a gap, represented by parameter $\Delta$, 
in the band structure such that $\mathcal{H}_{0}+\mathcal{H}_{\Delta}$ 
describes a gapped spectrum with spin-degenerate parabolic
CB and VB. In addition the bands are spin-split due to SOC which is captured by 
the term $\mathcal{H}_{\textrm{soc}}$ with the parameters $\lambda_{\textrm{c}}$ and $\lambda_{\textrm{v}}$ describing 
the spin splitting of the CB and VB.
The Hamiltonian $\mathcal{H}_{0}+\mathcal{H}_{\Delta}+\mathcal{H}_{\textrm{soc}}$ is already suitable to describe the spectrum of intrinsic TMDCs around the band edges at K and K'. 
In the case when we have a ferromagnetic substrate, proximity exchange effects are present
and we introduce the term $\mathcal{H}_{\textrm{ex}}$,
with $B_{\textrm{c}}$ and $B_{\textrm{v}}$ describing the proximity induced exchange splittings. 
Note that this term explicitly breaks time-reversal symmetry and thus the valley degeneracy. 
Finally, a Rashba term $\mathcal{H}_{\textrm{R}}$, with $\lambda_{\textrm{R}}$ being the Rashba parameter, can be present since a substrate breaks 
the inversion symmetry of the TMDC.

\section{Computational details}
\subsection{First-principles calculations}
\label{sec:Computational}
The electronic structure calculations and structural relaxation of 
our geometries are performed with density functional theory (DFT)~\cite{Hohenberg1964:PRB} 
using \textsc{Quantum Espresso}~\cite{Giannozzi2009:JPCM}.
Self-consistent calculations are performed with the $k$-point 
sampling of $18\times 18\times 1$ for the TMDC/hBN/FM heterostrucures.
We perform open shell calculations that provide the 
spin polarized ground state, when a FM substrate is present. 
We use an energy cutoff for charge density of $550$~Ry, and
the kinetic energy cutoff for wavefunctions is $65$~Ry for the scalar relativistic pseudopotential 
with the projector augmented wave method \cite{Kresse1999:PRB} with the 
Perdew-Burke-Ernzerhof exchange correlation functional \cite{Perdew1996:PRL}.
When SOC is included, the fully relativistic versions of the pseudopotentials are used. 
In addition we include the Hubbard correction for the FMs Co and Ni with $U = 1$~eV \cite{Liechtenstein1995:PRB}.
For this value of the Hubbard $U$, the calculated magnetic moments of Co and Ni in our slab geometry are 
comparable to the magnetic moments of bulk hcp Co ($1.7~\mu_{\textrm{B}}$) and fcc Ni ($0.6~\mu_{\textrm{B}}$) \cite{Tung2012:PRB, Mohammed2012:JMMM}, justifying our choice of
$U = 1$~eV.
For the relaxation of the heterostructures, we add 
van der Waals corrections \cite{Grimme2006:JCC,Barone2009:JCC} and use 
quasi-newton algorithm based on trust radius procedure. 
In order to simulate quasi-2D systems, a vacuum of at least $16$~\AA~is used 
to avoid interactions between periodic images in our slab geometries.
Dipole corrections \cite{Bengtsson1999:PRB} are included for heterostructure calculations 
to get correct band offsets and internal electric fields.
Structural relaxations are performed until all components of all forces 
were reduced below $10^{-3}$~[Ry/$a_0$], where $a_0$ is the Bohr radius.

\subsection{Absorption spectra calculations for excitons}

To compute the excitonic spectra, we employ the effective Bethe-Salpeter equation 
(BSE)\cite{Rohlfing2000:PRB,Qiu2013:PRL,Soklaski2014:APL,Scharf2017:PRL,Scharf2019:JPCM,Zollner2019:strain,Zollner2019a:PRB,FariaJunior2019:arxiv} 
focusing on direct intralayer excitons at zero temperature and without doping effects. 
The specific form of the BSE we use is given in the Supplemental Material of 
Ref.~[\onlinecite{Zollner2019a:PRB}]. The single-particle spectra is obtained using 
the model Hamiltonian of Eq.~(\ref{Eq:Hamiltonian}) fitted to the first-principles 
band structures. Because of the coupling between conduction and valence band states 
(due to the Fermi velocity term), the direct term of the electron-hole interaction 
is given by
\begin{equation}
\mathbb{D}_{c^{\prime}v^{\prime}\vec{k}^{\prime}}^{cv\vec{k}}=-\Delta_{c^{\prime}v^{\prime}\vec{k}^{\prime}}^{cv\vec{k}}\,\text{V}(\vec{k}-\vec{k}^{\prime}) \, ,
\label{eq:D}
\end{equation}
with $\text{V}(\vec{k}-\vec{k}^{\prime})$ being the electrostatic potential and the 
mixing term $\Delta_{c^{\prime}v^{\prime}\vec{k}^{\prime}}^{cv\vec{k}}$ can be written 
in a very general way as
\begin{equation}
\Delta_{c^{\prime}v^{\prime}\vec{k}^{\prime}}^{cv\vec{k}}\!=\!\left[\sum_{l}^{\textrm{M}}\beta{c,l}^{*}(\vec{k})\beta_{c^{\prime},l}(\vec{k}^{\prime})\right]\!\!\left[\sum_{m}^{\textrm{M}}\beta_{v^{\prime},m}^{*}(\vec{k}^{\prime})\beta_{v,m}(\vec{k})\right] \, ,
\label{eq:mixing}
\end{equation}
in which $\textrm{M}$ is the size of the effective Hamiltonian, $c,c^\prime$ ($v,v^\prime$) 
are the conduction (valence) bands, and $\beta_{n,l}(\vec{k})$ is the $l$-th coefficient 
($l=1,...,\textrm{M}$) of the band $n$ at point $\vec{k}$. We point out that for 
decoupled models, for instance parabolic dispersions described by effective masses, 
the mixing term (\ref{eq:mixing}) would not be present and the BSE is reduced to the Wannier equation in $k$-space\cite{Haug2009book}. 

We model the electron-hole electrostatic potential $\text{V}(\vec{k}-\vec{k}^{\prime})$ 
that appears in Eq.(\ref{eq:D}) by the widely used Rytova-Keldysh 
potential \cite{Rytova1967:MUPB,Keldysh1979:JETP,Cudazzo2011:PRB,Berkelbach2013:PRB,Chernikov2014:PRL,Wu2015:PRB}.
Such potential essentially describes the electrostatics of a thin slab, i. e. the 2D 
material, with different top and bottom dielectric environments. In $k$-space the 
Rytova-Keldysh potential is given by
\begin{equation}
\text{V}(\vec{K})=\frac{1}{\mathcal{A}}\frac{e^{2}}{2\varepsilon_{0}}\frac{1}{\varepsilon K+r_{0}K^{2}} \, ,
\end{equation}
with $\vec{K}=\vec{k}-\vec{k}^{\prime}$, $\mathcal{A}$ is the 2D unit area, $e$ is the 
electron charge, $\varepsilon_0$ is the vacuum permittivity, $r_0$ is the screening 
length of the 2D material and $\varepsilon$ is the effective dielectric constant given 
by the average of the top and bottom dielectric constants, i. e., 
$\varepsilon=\left(\varepsilon_{t}+\varepsilon_{b}\right)/2$. For the screening lengths 
of the TMDCs we used the values provided in Ref.~[\onlinecite{Berkelbach2013:PRB}]. 

The BSE is solved on a 2D $k$-grid from -0.5 to 0.5 
$\textrm{\AA}^{-1}$ in $k_x$ and $k_y$ directions with total discretization of 
$101 \times 101$ points (leading to a spacing of $\Delta k = 10^{-2} \; \textrm{\AA}^{-1}$). 
To improve convergence, the Coulomb potential is averaged around each $k$-point in a 
square region of $-\Delta k /2$ to $\Delta k /2$ discretized with $101 \times 101$ points\cite{Scharf2017:PRL,Zollner2019a:PRB,Zollner2019:strain}.

The absorption spectra for a photon with energy $\hbar\omega$, incorporating excitonic 
effects is given by\cite{Scharf2017:PRL,Scharf2019:JPCM}
\begin{equation}
\alpha^{a}(\hbar\omega)=\frac{4\pi^{2}e^{2}}{\varepsilon_{0}c_{l}\omega \mathcal{A}\hbar^{2}}\sum_{N}\left|\sum_{cv\vec{k}}A_{cv\vec{k}}(N)p_{vc}^{a}(\vec{k})\right|^{2}\!\!\delta\left(\Omega_{N}-\hbar\omega\right) \, ,
\end{equation}
with the superindex $a$ indicating the polarization of the light, $e$ is the electron 
charge, $\varepsilon_0$ is the vacuum permittivity, $c_l$ is the speed of light (the 
subindex $l$ was added to not be confused with the conduction band index $c$), 
$\mathcal{A}$ is the 2D unit area, the summation indices $c$ and $v$ label the 
conduction and valence bands, respectively, $\vec{k}$ is the wave vector, $N$ 
labels the excitonic states, $A_{cv\vec{k}}(N)$ is the exciton envelope function, 
$\Omega_N$ is the exciton energy and the single-particle dipole matrix  
element is $p_{nm}^{a}(\vec{k})=\frac{\hbar}{m_{0}}\left\langle  n,\vec{k}\left|\hat{e}_{a}\cdot\vec{p}\right|m,\vec{k}\right\rangle$ 
(computed using the eigenstates of the model Hamiltonian). Finally, to the 
calculated absorption spectra we apply a lorentzian broadening with energy 
dependent full width at half-maximum\cite{Haug2009book, Scharf2019:JPCM}
\begin{equation}
\Gamma(\hbar\omega)=\Gamma_{1}+\frac{\Gamma_{2}}{1+e^{\left[\left(E_{0}-\hbar\omega\right)/\Gamma_{3}\right]}}
\end{equation}
using $\Gamma_1=\Gamma_2=\Gamma_3=10\;\text{meV}$. $E_0$ is the single-particle energy at the K point for the first allowed optical transition.

\section{TMDC/hBN/FM heterostructures}
\subsection{Geometry}
\begin{figure}[htb]
 \includegraphics[width=.95\columnwidth]{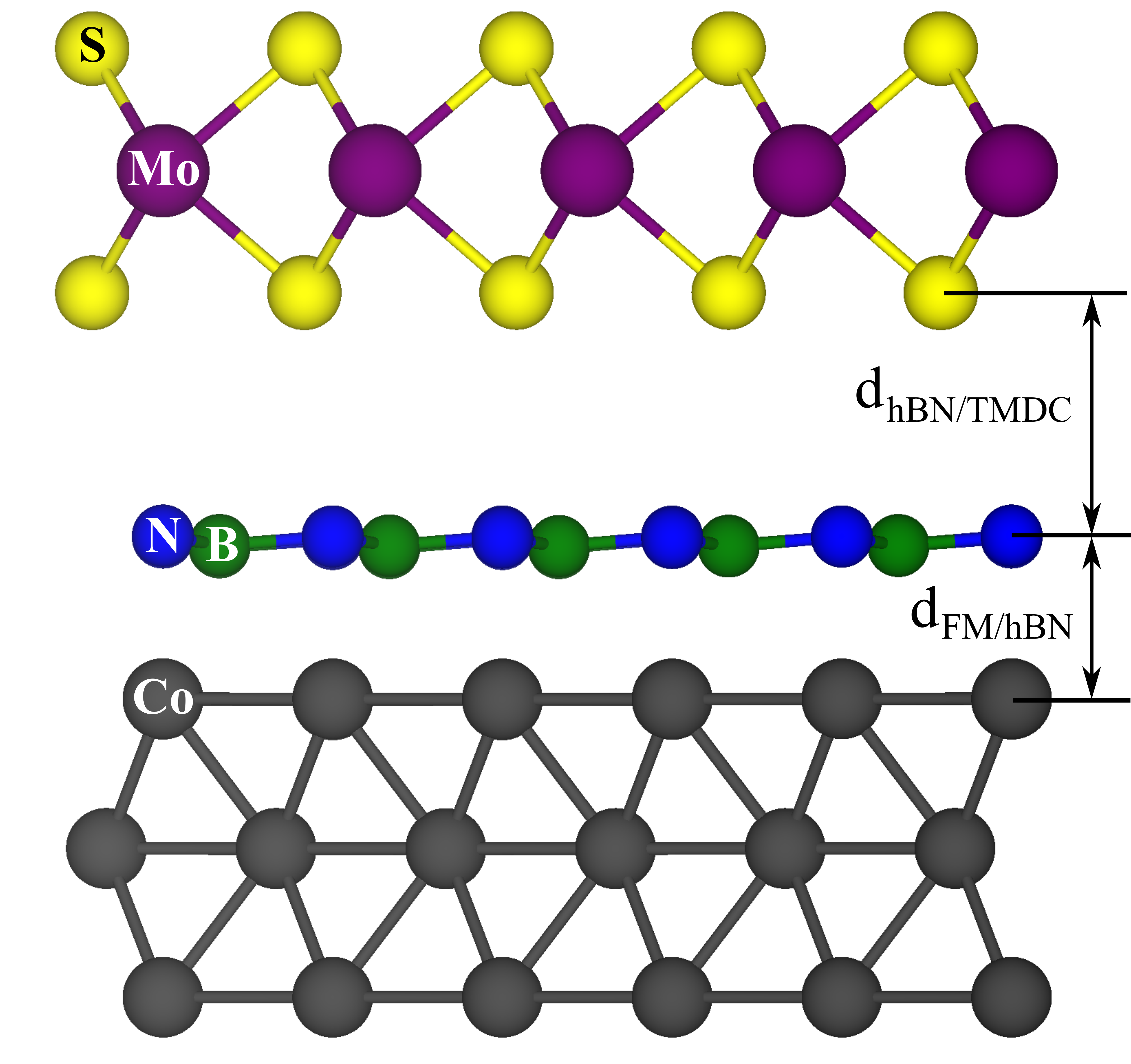}
 \caption{(Color online) Side view of the TMDC/hBN/FM structure with labels for the atoms and distances. 
 As an example, we show here MoS$_2$/hBN/Co. The distances between the layers $d_{\textrm{FM/hBN}}$ and 
 $d_{\textrm{hBN/TMDC}}$ are listed in Tab. \ref{Tab:fit_TMDC_hBN_FM} for all considered heterostructures. 
 }\label{Fig:structure_TMDC_hBN_FM}
\end{figure}

In order to calculate the proximity exchange in a TMDC/hBN/FM heterostructure, 
we have to find a common unit cell for all compounds minimizing strain effects. 
In Fig. \ref{Fig:structure_TMDC_hBN_FM} we show the geometry 
for MoS$_2$/hBN/Co, as an exemplary structure.
Initial atomic structures are set up with the atomic simulation environment (ASE) \cite{ASE}, as follows.
We choose a $4\times 4$ supercell of the TMDC (MoS$_2$, WS$_2$), 
a $5\times 5$ cell of hBN, and a $5\times 5$ cell of the FMs (Co, Ni). 
For the FM we take three monolayers of hcp Co
or fcc(111) Ni. The hBN is placed above the FM, such that the nitrogen 
sits above the topmost Co/Ni atom, and the
boron above the fcc-position above the FM slab, 
as found by previous studies \cite{Zollner2016:PRB}. 
The TMDC is placed such above the hBN/FM slab, that a transition metal atom
(Mo, W) sits above a nitrogen atom. 
By choosing the heterostructure as explained, our unit cell contains 173 atoms, 
with a lattice constant of $a = 12.637$~\AA, for all of our
considered hybrid geometries. For that we have to modify the 
lattice constants of the subsystem layers (TMDC, hBN, FM). 
In Tab. \ref{Tab:strain_subsystems} we give an overview on the 
original experimental and the new modified lattice constants used for the 
heterostructures, as well as the introduced strain.
We can see that a maximum strain of about 1.5\% is present for Ni, 
being still acceptable for our purposes.

\begin{table}[htb]
\begin{ruledtabular}
\begin{tabular}{lccccc}
 & Co & Ni & hBN & MoS$_2$ & WS$_2$\\
 \colrule
 $a$ (exp.) [\AA] & 2.507 & 2.492 & 2.504 & 3.150 & 3.153 \\
 $a$ (het.) [\AA] & 2.527 & 2.527 & 2.527 & 3.159 & 3.159 \\
 strain [\%] &	0.8 & 1.4 & 0.9	& 0.3 & 0.2
\end{tabular}  
\end{ruledtabular}
\caption{\label{Tab:strain_subsystems} Overview of the lattice constants and strains for the subsystems used in the 
TMDC/hBN/FM heterostructures.
The experimental $a$ (exp.) lattice constants 
(Refs. \onlinecite{Singal1977:PRB, Kittel2004, Catellani1987:PRB, Wakabayashi1975:PRB, Schutte1987:JSSC}) 
of the bulk systems and lattice 
constants used for the heterostructures $a$ (het.) are given, along with the
introduced strain for each subsystem, calculated as $(a_{\textrm{het}}-a_{\textrm{exp}})/{a_{\textrm{exp}}}$.
Note that for nickel, it is the lattice constant of a fcc(111) quasi-hexagonal surface. 
}
\end{table}

To determine the interlayer distances, the atoms of the TMDC, hBN, and the top two Co/Ni layer atoms
were allowed to relax only in their $z$ positions (vertical to the layers).
The average distances between the layers $d_{\textrm{FM/hBN}}$ and 
$d_{\textrm{hBN/TMDC}}$, as defined in Fig. \ref{Fig:structure_TMDC_hBN_FM}, 
are listed in Tab. \ref{Tab:fit_TMDC_hBN_FM} for all considered geometries.
The distances are measured from the average position of the top Co/Ni (bottom S) atoms,
with respect to the average position of the N atoms of the hBN layer.  
The corrugation of the hBN is on average $0.12$~\AA, independent of the heterostructure.
The distances $d_{\textrm{FM/hBN}}$ between the FMs and hBN are around $2.1$~\AA, indicating strong bonding.
The distances $d_{\textrm{hBN/TMDC}}$ between the hBN and the TMDCs are roughly $3.15$~\AA, 
being in the range of typical van der Waals distances.
All these observations are in agreement with previous calculations of hBN 
on metallic substrates \cite{Zollner2016:PRB, Bokdam2014:PRB, Lazic2014:PRB}, 
as well as TMDCs on hBN/metal interfaces \cite{Farmanbar2016:PRB,Farmanbar2015:PRB}.

In Tab. \ref{Tab:magn_moments} we summarize the calculated averaged atomic magnetic moments. For example, in the case of MoS$_2$/hBN/Co, we find that 
the average magnetic moments
in the first, second, and third Co-layer are $1.77, 1.71$, and $1.60~\mu_{\textrm{B}}$, ordered from vacuum to the hBN-interface. 
The FM metal layer induces a small negative (positive) magnetic moment in the B (N) atoms of about $-0.010~(0.003)~\mu_{\textrm{B}}$. 
In the case of Ni, the magnitude of proximity induced magnetic moments in the hBN layer are different compared to Co. Due to the different lattices and magnetic moments of fcc(111) Ni and hcp Co, the hybridization of FM $d$-orbitals and hBN $p$-orbitals and the resulting proximity exchange strength are not completely the same in the two cases. 
Finally, the Mo atoms have a magnetic moment of about $0.004~\mu_{\textrm{B}}$, which is responsible for the exchange splitting in the TMDC bands, that we analyze in the following. 

\begin{table}[htb]
\begin{ruledtabular}
\begin{tabular}{lccccc}
\multirow{3}{*}{system} & MoS$_2$ & MoS$_2$ & WS$_2$ & WS$_2$\\
 & hBN & hBN & hBN & hBN\\
 & Co & Ni & Co & Ni\\
 \colrule
Mo/W & 0.004 & 0.003 & 0.002 & 0.002\\
B & -0.010 & -0.004 & -0.010 & -0.004\\
N & 0.003 & 0.014 & 0.003 & 0.014\\
Co/Ni (3) & 1.60 & 0.61 & 1.60 & 0.61 \\
Co/Ni (2) & 1.71 & 0.76 & 1.71 & 0.76\\
Co/Ni (1) & 1.77 & 0.75 & 1.77 & 0.75 \\ 
\end{tabular}
\end{ruledtabular}
\caption{\label{Tab:magn_moments} Calculated atomic magnetic moments of 
the TMDC/hBN/FM systems given in $\mu_{\textrm{B}}$. The values are obtained by averaging the magnetic moments of the individual atomic species. For Co/Ni atoms we distinguish the three layers of the FM slab by numbers in brackets (1--3) ordered from vacuum to hBN-interface side.}
\end{table}

\subsection{Band structure analysis and fit results}
\begin{figure*}[htb]
 \includegraphics[width=.99\textwidth]{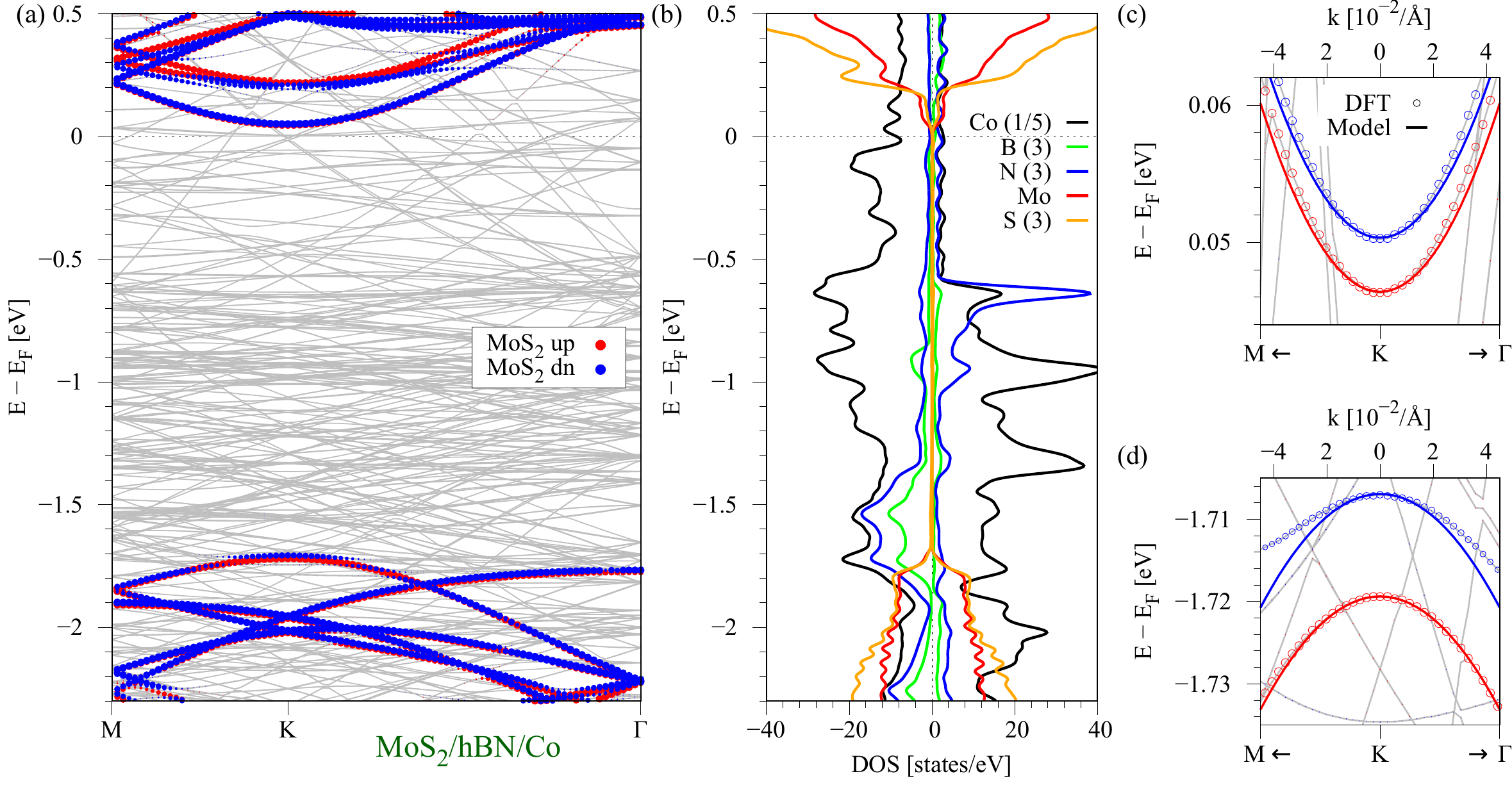}
 \caption{(Color online) Calculated band structure and density of states of the MoS$_2$/hBN/Co system without SOC. 
 (a) Band structure along high symmetry path M-K-$\Gamma$. The bands corresponding to MoS$_2$ are emphasized by red (spin up) and blue (spin down) spheres. Bands originating from the hBN/Co substrate are plotted in grey. 
 (b) The atom and spin resolved density of states of the heterostructure, where different colors correspond to different atoms. The number in brackets is 
 the multiplication factor, i. e., the contribution of, for example, B atoms is multiplied by a factor of 3. 
 (c) Zoom to the CB edge originating from MoS$_2$ around the K point. 
 Symbols are DFT data and solid lines are the fit to the model Hamiltonian. 
 (d) Same as (b), but for VB edge. 
 }\label{Fig:bands_EX_TMDC_hBN_Co}
\end{figure*}

In the following we analyze the results for MoS$_2$/hBN/Co, shown in Fig. \ref{Fig:bands_EX_TMDC_hBN_Co}, as a representative example. 
Similar to Fig. \ref{Fig:bands_EX_TMDC_hBN_Co}, 
we show the DFT results for the other TMDC/hBN/FM heterostructures in the Appendix. 

In Fig. \ref{Fig:bands_EX_TMDC_hBN_Co}(a) we show the calculated band structure without SOC of the MoS$_2$/hBN/Co heterostructure. Since a lot of bands are involved, originating from the hBN/FM substrate, we only show a projection of the bands originating from the TMDC. 
We find that the band structure closely resembles the bands of pristine MoS$_2$. 
The Fermi level is located below the TMDC CB edge.
Our calculated band structure is very different compared to other studies 
of proximity induced exchange \cite{Qi2015:PRB, Zhang2016:AM, Xu2018:PRB}, 
where the TMDC experiences strong doping, probably due to the polar surfaces 
of the substrates they consider.
Note that a polar surface does not reflect a realistic situation, 
as surface reconstructions are present in real experiments. 
This can strongly modify the proximity exchange 
in experiments, as the hybridization with the substrate is the main origin of proximity exchange. 
Such effects are absent in a hBN/FM hybrid substrate, due to the hBN buffer layer. 
Another advantage of our FM/hBN substrate is that the Curie temperature of the standard ferromagnets \cite{Kittel2004} Co (T$_{\textrm{C}}=1388$~K) and Ni (T$_{\textrm{C}}=627$~K) is well 
above room temperature, allowing for experiments at ambient conditions. 
The ultimate goal is the spin injection into the TMDC via the hBN/FM tunnel junction.

In Fig. \ref{Fig:bands_EX_TMDC_hBN_Co}(b) we show the atom and spin resolved density of states of the MoS$_2$/hBN/Co heterostructure. In the chosen energy window around the Fermi level, mainly the Co atoms contribute to the density of states. 
We also find that hBN is not an insulator anymore, as states from the N and B atoms are present in the whole energy range. Note that the actual band edges of the hBN do not 
reside within the shown energy window. 
Especially interesting is the relatively large contribution of N spin-up (spin-down) states at about $-0.65$ ($-1.7$)~eV below the Fermi level, which is an indication of strong hybridization with the FM layers. More precisely, the Co $d$ orbitals hybridize with the $p$ orbitals of the hBN layer and mediate the exchange coupling. 
The band edges of the MoS$_2$ layer can be also identified in the density of states at about $0$~eV and $-1.7$~eV. 
Especially near the VB edge of MoS$_2$ at around $-1.7$~eV, there is a significant contribution from B and N atoms in the spin-down channel. 
Thus, the Co $d$-orbitals hybridize, via the hBN buffer layer, 
with the VB spin-down states of MoS$_2$.

\begin{table}[htb]
\begin{ruledtabular}
\begin{tabular}{lccccc}
\multirow{3}{*}{system} & MoS$_2$ & MoS$_2$ & WS$_2$ & WS$_2$\\
 & hBN & hBN & hBN & hBN\\
 & Co & Ni & Co & Ni\\
 \colrule
 $\Delta$ [eV] & 1.761 & 1.769 & 1.910 & 1.910\\
 $v_{\textrm{F}}$ [$10^{5} \frac{\textrm{m}}{\textrm{s}}$] & 5.303 & 5.475 & 6.907 & 6.908 \\
 $B_{\textrm{c}}$ [meV]  & 1.964 &1.697  & 1.080 & 1.077\\
 $B_{\textrm{v}}$ [meV]  & 6.365 & 2.185 & 3.629 & 2.308\\
  dipole [Debye] & 1.940 & 2.871 & 2.386 & 3.713\\
 $d_{\textrm{FM/hBN}}$ [\AA] & 2.089 & 2.085 & 2.089 & 2.084\\
 $d_{\textrm{hBN/TMDC}}$ [\AA] & 3.157 & 3.217 &3.151 & 3.137
\end{tabular}
\end{ruledtabular}
\caption{\label{Tab:fit_TMDC_hBN_FM} Summary of the fit parameters, calculated dipoles and 
distances for TMDC/hBN/FM systems without SOC. 
The Hamiltonian used to fit these systems
is $\mathcal{H}_{0}+\mathcal{H}_{\Delta}+\mathcal{H}_{\textrm{ex}}$, with $\Delta$ as the orbital gap of the spectrum, the 
Fermi velocity $v_{\textrm{F}}$ and $B_{\textrm{c}}$ and $B_{\textrm{v}}$ are the proximity exchange parameters, respectively. 
The dipole of the structures is given in debye and  
$d_{\textrm{hBN/TMDC}}$ is the distance between hBN and the TMDC, and $d_{\textrm{FM/hBN}}$ is the distance between the FM and hBN,
as defined in Fig. \ref{Fig:structure_TMDC_hBN_FM}.
}
\end{table}

If we zoom to the fine structure around the K point, we find that the bands of 
the TMDC are spin split due to proximity exchange coupling, see Figs. \ref{Fig:bands_EX_TMDC_hBN_Co}(c,d). 
The splitting of the CB is smaller than that of the VB, both being in the
few meV range.
Since SOC effects are turned off, we can fit the band structure around the K point
to our model Hamiltonian, neglecting for now $\mathcal{H}_{\textrm{soc}}+\mathcal{H}_{\textrm{R}}$. 
In Figs. \ref{Fig:bands_EX_TMDC_hBN_Co}(c,d), we can see that the bands are nicely reproduced 
by the model with the fit parameters given in Tab. \ref{Tab:fit_TMDC_hBN_FM}.
However, we can see that there is a discrepancy between the model and the calculation, 
especially for the spin-down VB away from the K point. 
The case of MoS$_2$/hBN/Co is the only one, where this
happens. The origin of this, is the hybridization
of the VB spin-down states of MoS$_2$ with Co $d$-states, 
as one can see in Fig. \ref{Fig:bands_EX_TMDC_hBN_Co}(a,b).

In Tab. \ref{Tab:fit_TMDC_hBN_FM} we summarize our fit parameters for all considered heterostructures.
We notice, that especially the parameter 
$B_{\textrm{v}}$ for the MoS$_2$/hBN/Co case is very large, which is 
caused by the previously explained effect of hybridization. 
In a similar way, an earlier study has shown, that the hybridization with Co $d$-orbitals 
can strongly enhance the proximity exchange splitting in graphene on a hBN/Co substrate \cite{Zollner2016:PRB}. 
Note that the value of the Hubbard $U$, which shifts the Co $d$-levels in energy, affects the hybridization and the spin splitting, as shown in Ref. \onlinecite{Zollner2016:PRB}. 
Since we have considered $U = 1$~eV only, we can only predict that such a hybridization 
is present in experiments, which leads to an enhanced spin splitting.

Unfortunately, we were not able to proper converge 
the calculations of the TMDC/hBN/FM heterostructures including SOC effects.
The spin-orbit splitting in the TMDC VB is large \cite{Zollner2019:strain, Kormanyos2014:2DM} and may shift the corresponding bands to much in energy to spoil the hybridization. 
Therefore, we cannot be completely sure, whether the mentioned hybridization effect
in the MoS$_2$/hBN/Co structure will remain. 
However, a similar work of proximity exchange in TMDC/CrI$_3$ heterostructures has shown, 
that inclusion of SOC barely affects the proximity exchange parameters \cite{Zollner2019a:PRB}. 
Thus, we believe that proximity exchange on the order of 1--5~meV 
would still be present when including SOC for our TMDC/hBN/FM structures.

Similar to graphene/hBN/FM stacks \cite{Zollner2016:PRB}, 
we expect that the proximity exchange in the TMDCs also decreases with 
increasing number of hBN layers between the FM and the TMDC. 
Unfortunately, due to computational limitations (number of atoms in the supercell), 
we cannot study more than one hBN layer.

\subsection{TMDC/hBN subsystems}

Experimentally it is also interesting to consider the bare 
MoS$_2$/hBN and WS$_2$/hBN heterostructures, without any influence from the FM.
For that, we take the TMDC/hBN/FM geometries, but remove the FM layers.
\begin{table}[htb]
\begin{ruledtabular}
\begin{tabular}{lcc}
\multirow{2}{*}{system} & MoS$_2$ & WS$_2$ \\
 & hBN & hBN  \\
 \colrule
 $\Delta$ [eV] & 1.756 & 1.872  \\
 $v_{\textrm{F}}$ [$10^{5} \frac{\textrm{m}}{\textrm{s}}$] & 5.432 & 6.786  \\
 $\lambda_{\textrm{c}}$ [meV]  & -1.361 & 17.03 \\
 $\lambda_{\textrm{v}}$ [meV]  & 72.96 & 208.6  \\
 dipole [Debye] &-0.701 & -0.703  \\
  $d_{\textrm{hBN/TMDC}}$ [\AA] & 3.330 & 3.252 \\
\end{tabular}
\end{ruledtabular}
\caption{\label{Tab:fit_TMDC_hBN} Summary of the fit parameters, calculated dipoles and 
distances for TMDC/hBN systems with SOC. The Hamiltonian used to fit these systems is 
$\mathcal{H}_{0}+\mathcal{H}_{\Delta}+\mathcal{H}_{\textrm{soc}}+\mathcal{H}_{\textrm{R}}$, 
with $\Delta$ as the orbital gap of the spectrum, the Fermi velocity $v_{\textrm{F}}$ and $\lambda_{\textrm{c}}$ and $\lambda_{\textrm{v}}$ are the SOC parameters. 
The distances are given as defined in Fig. \ref{Fig:structure_TMDC_hBN_FM}.}
\end{table}
After subsequent relaxation, similar as described above, we calculate the electronic 
band structures for the \textit{subsystems}, including SOC effects. The corrugation of the hBN reduces to about 20~pm on average, 
and the distances $d_{\textrm{hBN/TMDC}}$ between hBN and the TMDC are given in Tab. \ref{Tab:fit_TMDC_hBN}
for the TMDC/hBN stacks. The FM is responsible for the corrugation of the hBN layer, as already pointed out in an earlier work \cite{Zollner2016:PRB}. 
Our model Hamiltonian is also suitable to describe this situation, when neglecting $\mathcal{H}_{\textrm{ex}}$.

\begin{figure}[htb]
 \includegraphics[width=.99\columnwidth]{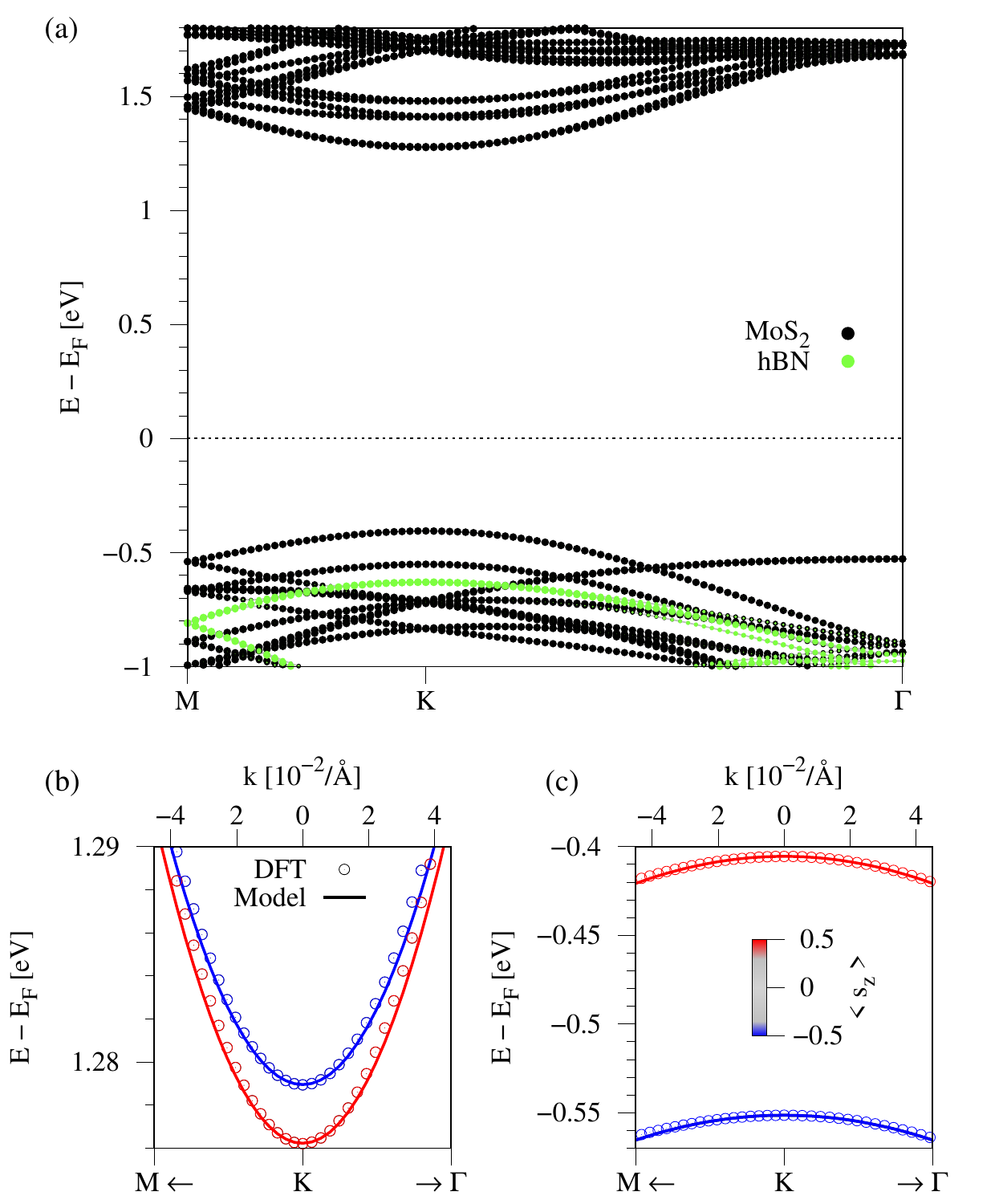}
 \caption{(Color online) Calculated band structure of MoS$_2$/hBN including SOC effects. 
 (a) Band structure along high symmetry lines. The bands corresponding to MoS$_2$ (hBN) are
 plotted in black (green). (b) Zoom to the CB edge. The color corresponds to the
 $s_z$-expectation value. Symbols are DFT data and solid lines are fits to the Model Hamiltonian. 
 (c) Same as (b), but for VB edge. 
 }\label{Fig:bands_SOC_TMDC_hBN}
\end{figure}

The calculated band structure for MoS$_2$/hBN is shown in Fig. \ref{Fig:bands_SOC_TMDC_hBN}.
We find that we can perfectly reproduce the band
structure around K and K' valleys, with the fit parameters given in Tab. \ref{Tab:fit_TMDC_hBN}.
The fit parameters for the TMDC/hBN heterostructures are nearly identical
to the ones obtained for the bare TMDC monolayers \cite{Zollner2019:strain, Kormanyos2014:2DM}.
\textit{We conclude that the hBN has effectively no impact on the TMDC dispersion and SOC.}
In principle, one would also expect Rashba SOC due to inversion symmetry breaking,
but from previous calculations of graphene/hBN/FM structures \cite{Zollner2016:PRB}, we conclude that the
proximity induced SOC due to the hBN/FM substrate is negligible, compared to the proximity induced exchange and the giant intrinsic SOC of the bare TMDCs \cite{Zollner2019:strain, Kormanyos2014:2DM}. 
Indeed, we find by fitting the model Hamiltonian to the band structure of the TMDC/hBN systems,
that $\mathcal{H}_{\textrm{R}}$ can be neglected, as the $s_z$ spin expectation values of 
the bands near the K and K' points almost do not differ from $\pm 0.5$, 
as we can see in Fig. \ref{Fig:bands_SOC_TMDC_hBN}.

In conclusion, the extracted proximity exchange (intrinsic SOC) parameters from the
calculations of the TMDC/hBN/FM (TMDC/hBN) systems, together with the Hamiltonian, 
can be used for further studies.

\section{Proximity exchange induced valley splitting}
The individually extracted parameters for proximity exchange and SOC, 
see Tabs. \ref{Tab:fit_TMDC_hBN_FM} and \ref{Tab:fit_TMDC_hBN}, in combination with our model
Hamiltonian can be used to calculate the low energy bands around K and K' point with both effects present. 
We average the parameters for $v_{\textrm{F}}$ and $\Delta$ from Tabs. \ref{Tab:fit_TMDC_hBN_FM} and \ref{Tab:fit_TMDC_hBN}, for the MoS$_2$ and WS$_2$ based structures. 
We take the SOC parameters for the two TMDCs from Tab. \ref{Tab:fit_TMDC_hBN}, and the proximity exchange parameters from Tab. \ref{Tab:fit_TMDC_hBN_FM}. 
The full parameter sets are summarized in Tab. \ref{Tab:exciton}, which we use for the following absorption spectra calculations. 

\begin{table}[htb]
\begin{ruledtabular}
\begin{tabular}{lccccc}
\multirow{3}{*}{system} & MoS$_2$ & MoS$_2$ & WS$_2$ & WS$_2$\\
 & hBN & hBN & hBN & hBN\\
 & Co & Ni & Co & Ni\\
 \colrule
 $\Delta$ [eV] & 1.759 & 1.763 & 1.891 & 1.891\\
 $v_{\textrm{F}}$ [$10^{5} \frac{\textrm{m}}{\textrm{s}}$] & 5.368 & 5.454 & 6.847 & 6.847 \\
  $\lambda_{\textrm{c}}$ [meV]  & -1.361 & -1.361 & 17.03 & 17.03\\
 $\lambda_{\textrm{v}}$ [meV]  & 72.96 & 72.96 & 208.6 & 208.6 \\
 $B_{\textrm{c}}$ [meV]  & 1.964 &1.697  & 1.080 & 1.077\\
 $B_{\textrm{v}}$ [meV]  & 6.365 & 2.185 & 3.629 & 2.308\\
 s. p. at K,K' [meV] & 8.80 & 0.98 & 5.10 & 2.46 \\
 $\textrm{A}_\textrm{1s}$ ($\varepsilon=2.75$) [meV] & 8.20 & 0.91 & 4.76 & 2.30 \\ 
 $\textrm{A}_\textrm{1s}$ ($\varepsilon=4.5$) [meV] & 8.31 & 0.93 & 4.84 & 2.34
\end{tabular}
\end{ruledtabular}
\caption{\label{Tab:exciton} Summary of the model Hamiltonian parameters, combined from Tabs. \ref{Tab:fit_TMDC_hBN_FM} and \ref{Tab:fit_TMDC_hBN}.
The orbital gap parameter $\Delta$, the 
Fermi velocity $v_{\textrm{F}}$, $B_{\textrm{c}}$ and $B_{\textrm{v}}$ are the proximity exchange parameters, and $\lambda_{\textrm{c}}$ and $\lambda_{\textrm{v}}$ are the SOC parameters. The valley splitting calculated from the single particle picture (s. part.) and 
from the absorption spectra of the $\textrm{A}_\textrm{1s}$ exciton peak. 
}
\end{table}

\begin{figure}[htb]
 \includegraphics[width=.99\columnwidth]{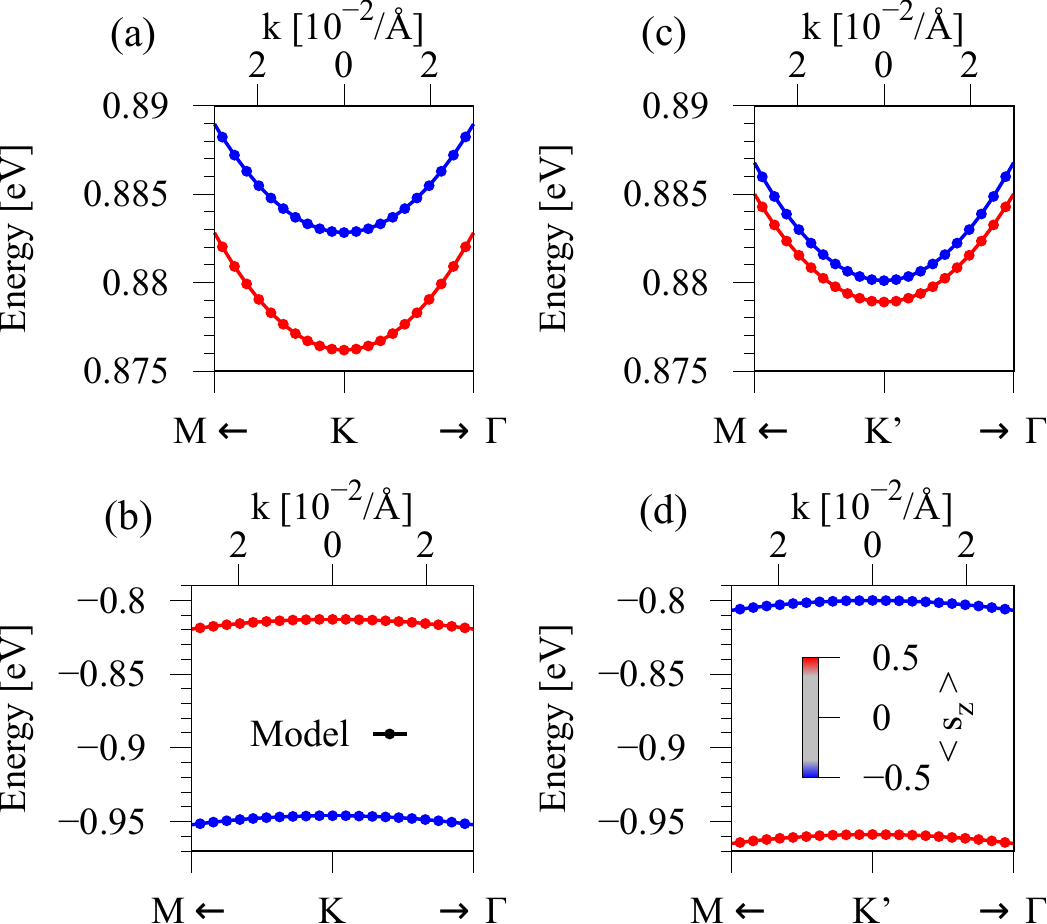}
 \caption{(Color online) Calculated model band structure employing the full Hamiltonian, for
 MoS$_2$/hBN/Co heterostructure using parameters from Tab. \ref{Tab:exciton}.
 (a,b) Low energy CB and VB at the K point. (c,d) Same as (a,b) but for the K' point. 
 }\label{Fig:bands_model}
\end{figure}

In this context, one can also generalize the exchange Hamiltonian \cite{Qi2015:PRB} to 
$\mathcal{H}_{\textrm{ex}}~=~-\hat{\textbf{m}}\cdot\mathbf{s} \otimes 
(B_{\textrm{c}}\sigma_{+} + B_{\textrm{v}} \sigma_{-})$, with 
$\hat{\textbf{m}}$ being a unit vector for the direction of the proximity exchange field 
and $\mathbf{s}$ is the vector containing Pauli spin matrices.
In Fig. \ref{Fig:bands_model} we show the calculated model band 
structure employing the full Hamiltonian with SOC and proximity 
exchange, setting $\hat{\textbf{m}} = \hat{m}_z$, for the MoS$_2$/hBN/Co heterostructure, using the parameters from Tab. \ref{Tab:exciton}. 
Due to the combination of SOC and proximity exchange, time-reversal symmetry is broken and
the valley degeneracy is lifted, as can be seen when comparing the spin-split CB at K and K', see Figs. \ref{Fig:bands_model}(a,c). 
Remarkably, the order of the spin bands in the CB is the same for the two valleys. 
For bare TMDCs, the spin splitting in the CB is determined by 
the corresponding SOC parameter $\lambda_{\textrm{c}}$. 
For the case of MoS$_2$, we find that the SOC parameter is comparable in magnitude with the
proximity exchange parameter $\lambda_{\textrm{c}} \approx B_{\textrm{c}}$, due to the hBN/FM
substrate. When the proximity exchange is larger than the SOC, the band ordering of the
spin-split CB is the same for both valleys, 
as can be seen in Figs. \ref{Fig:bands_model}(a,c). 
In contrast, the VB SOC parameter is much larger than the corresponding proximity exchange parameter $\lambda_{\textrm{v}} \gg B_{\textrm{v}}$, and the VB splitting is dominated by SOC.

\begin{figure}[htb]
 \includegraphics[width=.99\columnwidth]{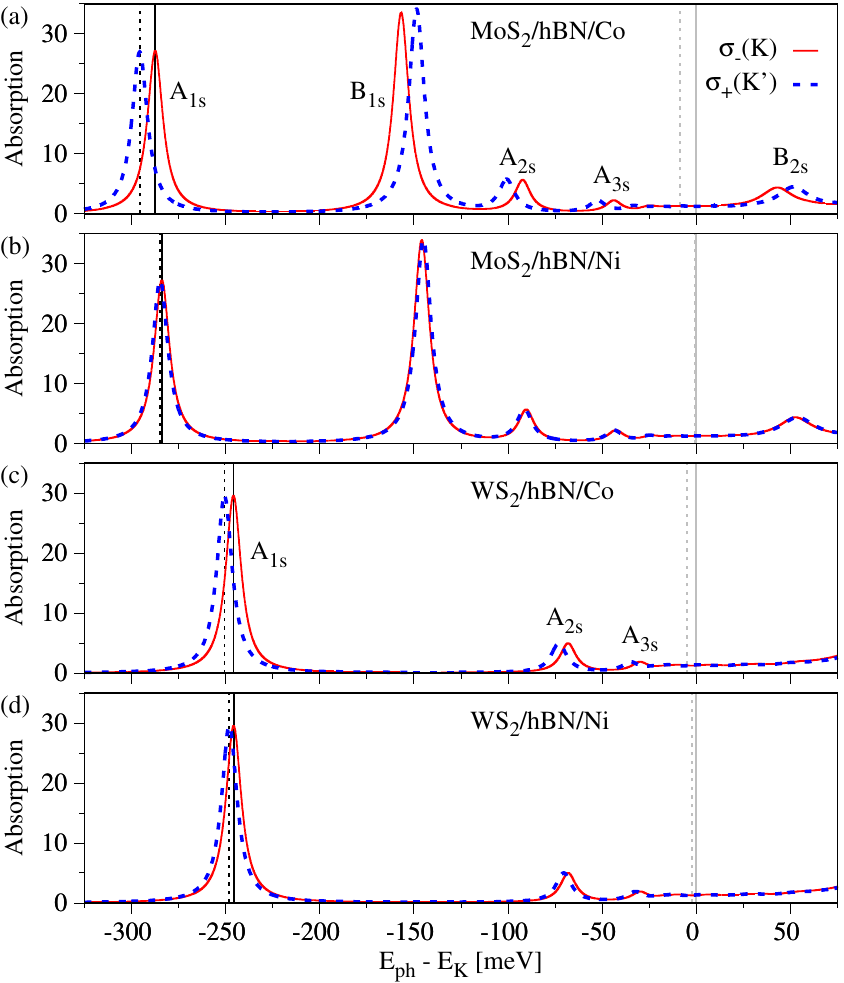}
 \caption{(Color online) Calculated absorption spectra for (a) MoS$_2$/hBN/Co, 
 (b) MoS$_2$/hBN/Ni, (c) WS$_2$/hBN/Co and (d) WS$_2$/hBN/Ni. In the x-axis, 
 $E_\textrm{ph}$ is the photon energy ($\hbar\omega$) and $E_\text{K}$ is the 
 optical energy  gap at the K point. The vertical solid (dashed) lines indicate the 
 energy contribution at the K (K') point. The vertical lines close to $E_\text{ph}-E_\text{K}=0$ indicate the single-particle energies while the 
 vertical lines at the $\textrm{A}_\textrm{1s}$ exciton peaks indicate the 
 excitonic contribution. The energy difference between $\sigma_+$ and $\sigma_-$ 
 polarizations is the valley splitting, summarized in Table.~\ref{Tab:exciton}. 
 Additional exciton peaks are also labeled. For these calculations we considered 
 an effective dielectric constant of $\varepsilon=2.75$.}
 \label{Fig:absorption}
\end{figure}

\begin{figure}[htb]
 \includegraphics[width=.99\columnwidth]{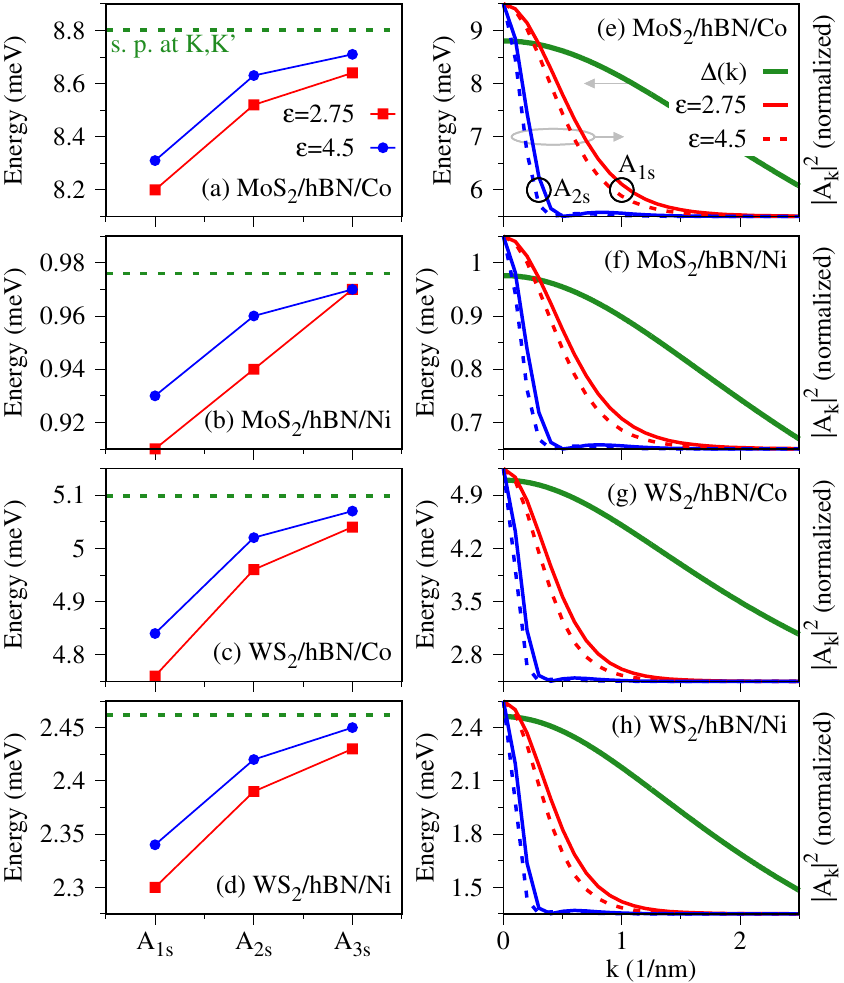}
 \caption{(Color online) Valley splitting of the $\textrm{A}_\textrm{1s}$, 
 $\textrm{A}_\textrm{2s}$ and $\textrm{A}_\textrm{3s}$ excitons in distinct 
 dielectric surroundings, $\varepsilon=2.75$ and $\varepsilon=4.5$, for 
 (a) MoS$_2$/hBN/Co, (b) MoS$_2$/hBN/Ni, (c) WS$_2$/hBN/Co and (d) WS$_2$/hBN/Ni. 
 The horizontal dashed line is the single-particle valley splitting directly at 
 K,K' (with values given in Table~\ref{Tab:exciton}). The valley splitting for 
 different excitons and dielectric surroundings can be understood by looking at 
 the spreading in momentum space of the single-particle valley splitting 
 $\Delta(\vec{k}) = E_{\textrm{K}} (\vec{k}) - E_{\textrm{K'}}(\vec{k})$ 
 (y-axis on the left side of the plot) and the exciton wavefunction (y-axis on 
 the right side of the plot)  for (e) MoS$_2$/hBN/Co, (f) MoS$_2$/hBN/Ni, 
 (g) WS$_2$/hBN/Co and (h) WS$_2$/hBN/Ni. In the x-axis, $k=0$ is at the K,K' 
 points. The exciton wavefunctions are scaled to have the same value at $k=0$.}
 \label{Fig:exciton_valley_splitting}
\end{figure}

For a more realistic assessment of the proximity exchange in the optical spectra, 
we investigate the valley splitting of the excitonic levels, employing the 
parameters summarized in Tab.~\ref{Tab:exciton} for the model Hamiltonian and the 
effective BSE. Let us first look at the absorption spectra assuming an effective 
dielectric constant of $\varepsilon=2.75$ as a model case of a TMDC placed on 
hBN ($\varepsilon_b=4.5$\cite{Stier2018:PRL}) and vacuum on top 
($\varepsilon_t=1$), as suggested by the geometry shown in 
Fig.~\ref{Fig:structure_TMDC_hBN_FM}. The calculated absorption spectra for the 
TMDC intralayer excitons is shown in Fig.~\ref{Fig:absorption} as function of 
the photon energy, $E_{\textrm{ph}}$, relative to the optical energy gap at the K point,
$E_\textrm{K}$ (the energy of the first allowed optical transition). With this x-axis 
convention, the energy of the first absorption peak, is roughly the $\textrm{A}_\textrm{1s}$ 
exciton binding energy. Because we are assuming a TMDC embedded in a dielectric 
environment with effective $\varepsilon=2.75$, the exciton binding energy is 
smaller than in the freestanding case (with $\varepsilon=1$ the binding energies 
are typically 0.5-0.6 eV for TMDCs \cite{Berkelbach2013:PRB,Jiang2017:PRL}). 
Due to the induced proximity 
exchange, K and K' valleys have now different transition energies and this is 
directly reflected in the absorption spectra as a splitting of the excitation 
energies with $\sigma_+$ and $\sigma_-$ polarizations. The extracted values for 
the valley splitting of the $\textrm{A}_\textrm{1s}$ exciton peak and for the 
single-particle picture at the K,K' points are summarized in Tab.~\ref{Tab:exciton}. 
Along with the visible excited excitonic states $\textrm{A}_\textrm{2s}$ and 
$\textrm{A}_\textrm{3s}$, the B excitons are also 
visible for the MoS$_2$-based systems. For instance, $\textrm{B}_\textrm{1s}$ excitons show large peaks around 
-150 meV and have a valley splitting with the same value as the 
$\textrm{A}_\textrm{1s}$ exciton but with opposite sign. 

What about the valley splitting of other exciton peaks and the effect of stronger 
dielectric surroundings? In Figs.~\ref{Fig:exciton_valley_splitting}(a)-(d) we show 
the valley splittings of the A-excitons (1s, 2s, and 3s) visible in 
Fig.~\ref{Fig:absorption} for effective dielectric constants of $\varepsilon=2.75$ 
and $\varepsilon=4.5$ (resembling a case for hBN encapsulated TMDC or assuming 
a stronger screening due to the presence of Co/Ni below the hBN layer\cite{Raja2017:NC}). 
Essentially, 
excited excitonic states and larger values of $\varepsilon$ provide larger valley 
splittings which tend to converge to the single-particle values obtained directly at 
the K,K' points. In order to grasp the physics behind this behaviour, we can look 
at the expectation value of the exciton valley splitting given by $\left\langle A^{\textrm{K}}_{\vec{k}} \left| E_{\textrm{K}} (\vec{k}) \right| A^{\textrm{K}}_{\vec{k}} \right\rangle - \left\langle A^{\textrm{K'}}_{\vec{k}} \left| E_{\textrm{K'}} (\vec{k}) \right| A^{\textrm{K'}}_{\vec{k}} \right\rangle$. Since the exciton wavefunctions are 
nearly identical for K and K' points (due to same effective masses) we can simply write 
the expectation value of the valley splitting as $\left\langle A_{\vec{k}} \left| \Delta(\vec{k}) \right| A_{\vec{k}} \right\rangle$, 
in which $\Delta(\vec{k}) = E_{\textrm{K}} (\vec{k}) - E_{\textrm{K'}}(\vec{k})$ is 
the single-particle valley splitting in momentum space. The spreading of $\Delta(\vec{k})$ 
(calculated with the model Hamiltonian) and the exciton envelope functions are shown in 
Figs.~\ref{Fig:exciton_valley_splitting}(e)-(h). Since $\Delta(\vec{k})$ has a maximum 
value at $\vec{k}=0$, the largest exciton valley splitting in this case is achieved 
for excitons more localized at $\vec{k}=0$, which is precisely the effect of looking at 
excited exciton states or increasing the effective dielectric constant as shown in 
Figs.~\ref{Fig:exciton_valley_splitting}(a)-(d). This intuitive picture of looking at 
the expectation value within the exciton framework has been recently employed to 
successfully explain the evolution of the g-factor in excited exciton 
states\cite{Chen2019:NL}. Furthermore, our analysis suggests that one possibility to 
engineer larger exciton valley splittings is to look for proximity exchange
effects that increase as one moves away of the K,K' edges.

As a final remark, we point out that the optical excitation energy difference 
between K and K' valleys for MoS$_2$/hBN/Co is giant, around 8~meV, 
translating into about 50~T magnetic field for bare TMDCs, if assuming 
0.15~meV/T valley splitting\cite{Srivastava2015:NP, Aivazian2015:NP, Li2014:PRL, MacNeill2015:PRL}. 
A similar giant valley splitting of about 5~meV is achieved in the case of 
WS$_2$/hBN/Co, translating to about 30~T magnetic field. Remarkably, in the case 
of MoS$_2$/hBN/Co, we have seen that proximity exchange together with band 
hybridization to $d$-orbitals causes a large VB splitting, leading to the giant 
valley splitting. In the case of WS$_2$/hBN/Co, the valley splitting is also giant, 
but without any band hybridization effects. When Ni is considered as the FM, we 
find smaller valley splittings, $\sim$1~meV (7~T) for MoS$_2$/hBN/Ni and 2.3~meV 
(15~T) for WS$_2$/hBN/Ni, but still the corresponding magnetic fields are quite 
large, compared to the valley splittings achieved without proximity exchange.  
Furthermore, our exciton calculations show that the valley splitting 
increases for excited exciton states, which could be observed experimentally in reflectance/transmittance measurements\cite{Chernikov2014:PRL,Stier2018:PRL}. 
Besides a direct observation of the exciton valley splitting, recent studies 
have shown that the induced proximity exchange has visible signatures in the 
magneto-optical Kerr response of excitons\cite{Catarina2020:2DM,Henriques2020:PRB}.

\section{Summary}
We have calculated the band structures of TMDC/hBN/FM heterostructures
and extracted valuable proximity exchange and SOC parameters, 
using a minimal low energy Hamiltonian. 
Due to the hBN buffer layer, the TMDC preserves a great degree of autonomy of its electronic structure.
Proximity exchange is found to be on the order of 1--10 meV, depending on the specific FM. 
Especially in MoS$_2$/hBN/Co, the spin splitting is giant, about 10~meV, due to proximity exchange and hybridization of Co $d$ states with the spin-down VB of the TMDC.
The excitonic absorption spectra, shows a giant valley splitting for the exciton peaks, translating into a valley polarization corresponding to tens of Tesla exchange field for bare TMDC monolayers.
The Ni-based heterostructures show less strong proximity exchange. We believe our calculations provide useful insight to interpret experimental properties of 
TMDC/hBN/(Co, Ni) devices, for instance related to spin injection, spin tunneling, and optics. Finally, our extracted parameters can be used as input for transport simulations and additional studies of excitonic effects.

\acknowledgments
We thank A. Chernikov for helpful discussions. This work was supported by the Deutsche
Forschungsgemeinschaft (DFG, German Research Foundation) SFB 1277 (Project-ID 314695032), 
DFG SPP 1666, the European Unions Horizon 2020 research and innovation program (Grant No. 785219), and the Alexander von Humboldt Foundation and Capes (grant No. 99999.000420/2016-06).

\appendix*
\section{Further DFT results}

\begin{figure*}[!htb]
 \includegraphics[width=.99\textwidth]{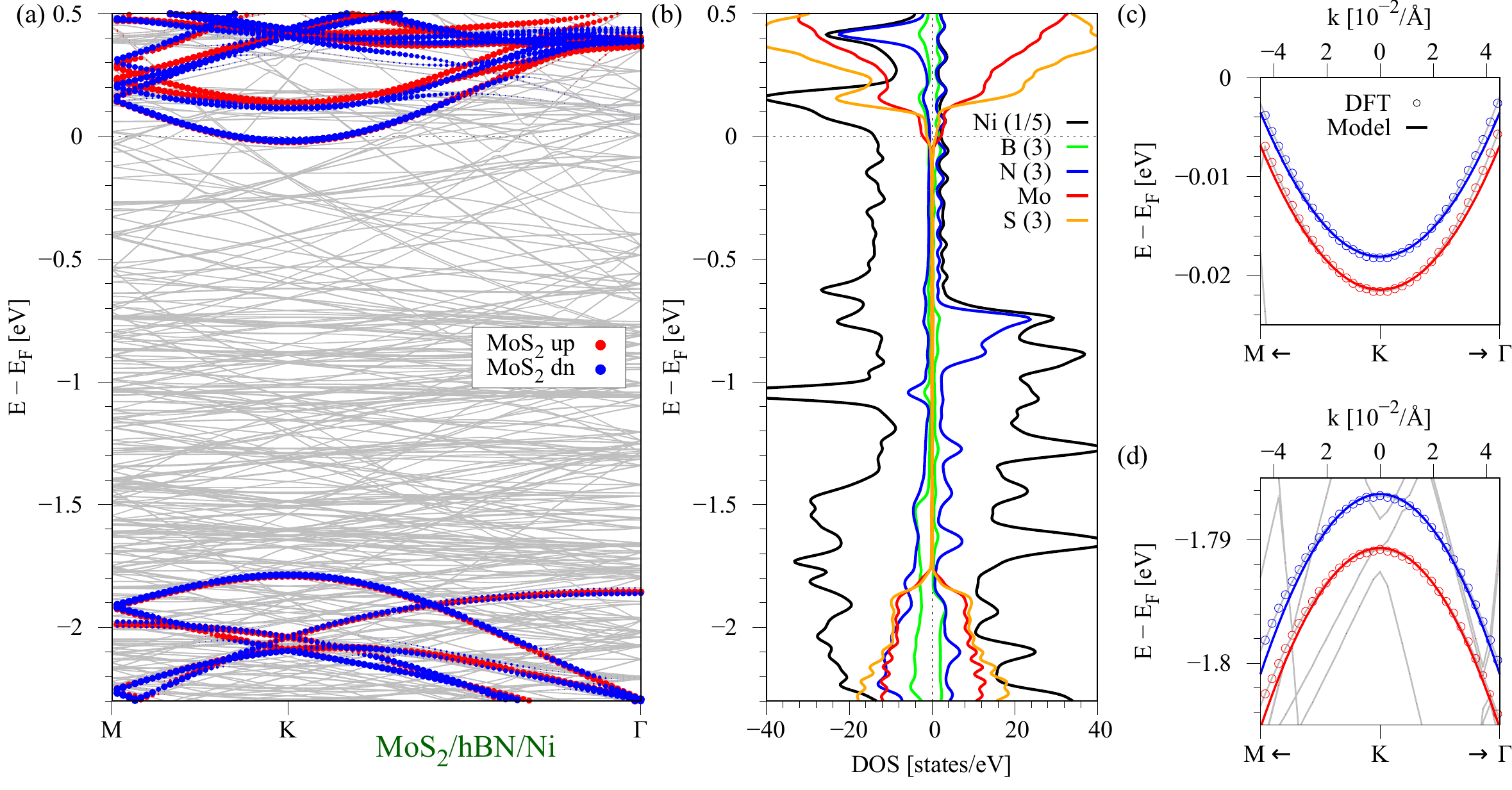}
 \caption{(Color online) Same as Fig. \ref{Fig:bands_EX_TMDC_hBN_Co}, but for the MoS$_2$/hBN/Ni system. 
 }\label{Fig:MoS2_hBN_Ni}
\end{figure*}

\begin{figure*}[!htb]
 \includegraphics[width=.99\textwidth]{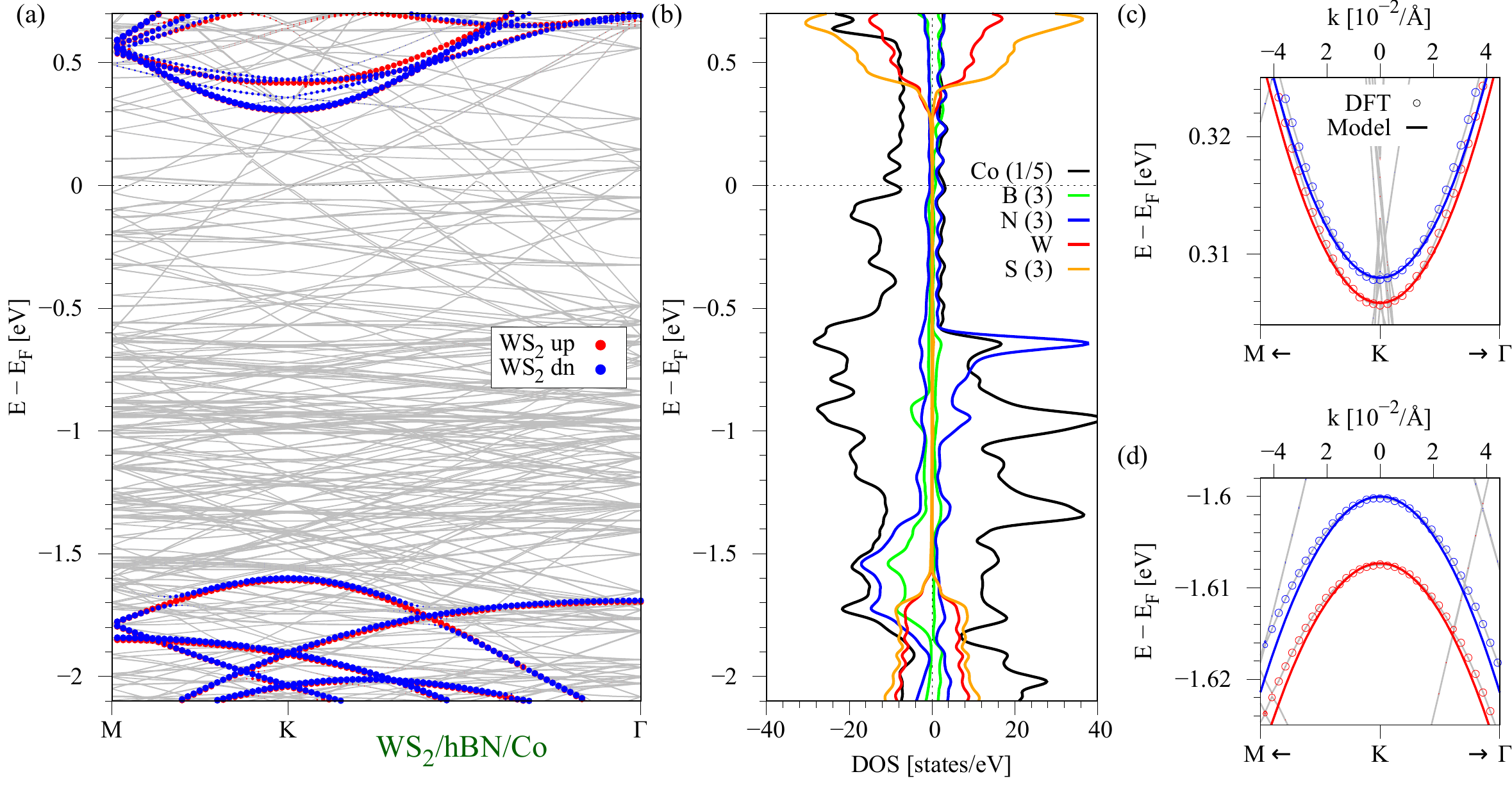}
 \caption{(Color online) Same as Fig. \ref{Fig:bands_EX_TMDC_hBN_Co}, but for the WS$_2$/hBN/Co system. 
 }\label{Fig:WS2_hBN_Co}
\end{figure*}

\begin{figure*}[!htb]
 \includegraphics[width=.99\textwidth]{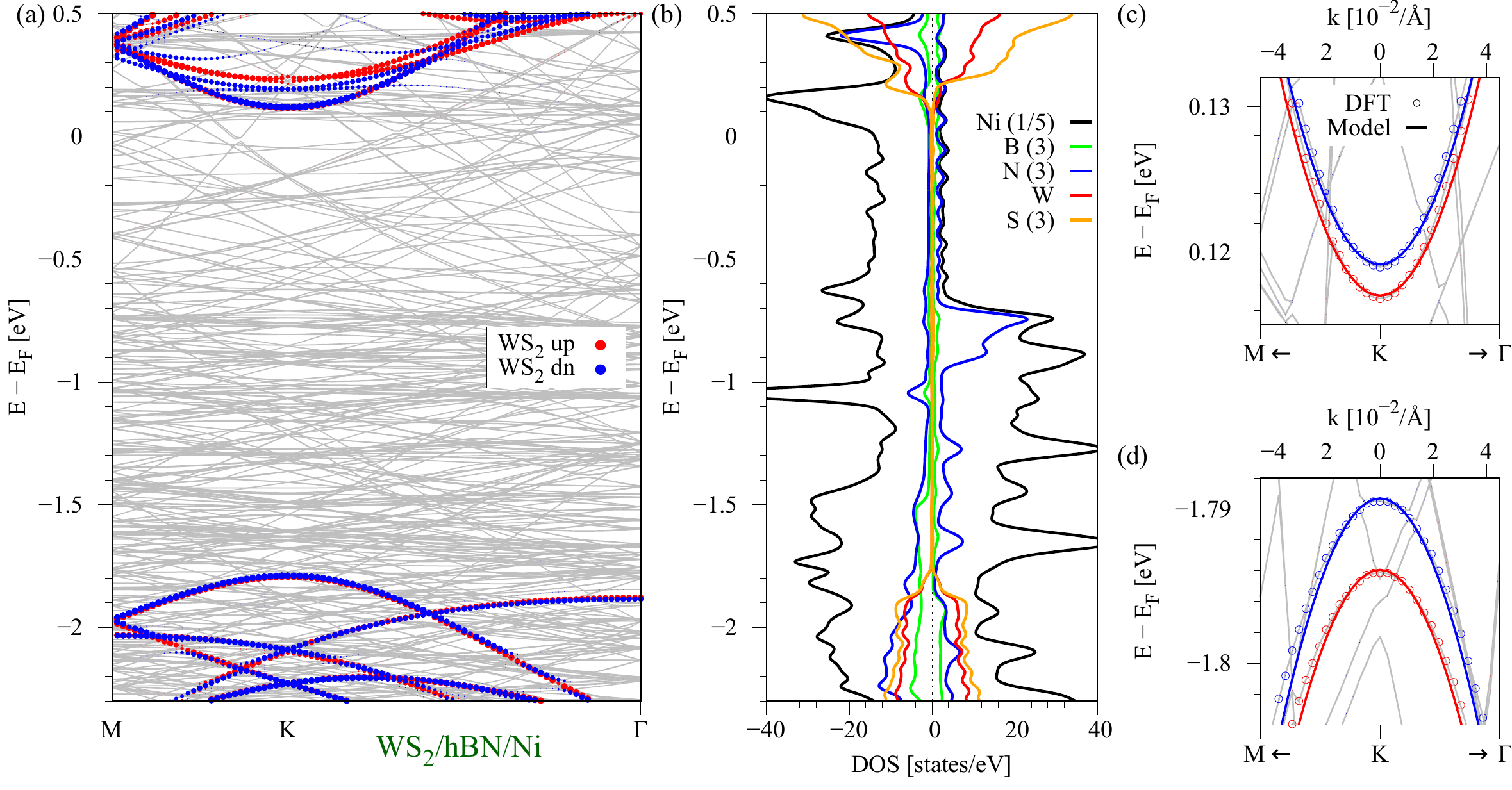}
 \caption{(Color online) Same as Fig. \ref{Fig:bands_EX_TMDC_hBN_Co}, but for the WS$_2$/hBN/Ni system. 
 }\label{Fig:WS2_hBN_Ni}
\end{figure*}

For completeness, in Fig. \ref{Fig:MoS2_hBN_Ni}, Fig. \ref{Fig:WS2_hBN_Co}, and Fig. \ref{Fig:WS2_hBN_Ni} we show the calculated band structure, density of states, and low energy model fits for the MoS$_2$/hBN/Ni, WS$_2$/hBN/Co, and WS$_2$/hBN/Ni heterostructures.


\bibliography{paper}
\end{document}